\documentclass[11pt]{article}
\usepackage{amsmath,amssymb,amsfonts}
\usepackage{latexsym}   
\begin{document}
\begin{titlepage}
\begin{flushright}
CP3-07-18\\
ICMPA-MPA/2007/17\\
June 2007\\
\end{flushright}
\begin{centering}
 
{\ }\vspace{0.5cm}
 
{\Large\bf The Weyl-Heisenberg Group on the Noncommutative Two-Torus:}

\vspace{5pt}

{\Large\bf A Zoo of Representations}

\vspace{1.8cm}

Jan Govaerts$^{\dagger,\ddagger,}$\footnote{Fellow of the Stellenbosch Institute
for Advanced Study (STIAS), Stellenbosch, Republic of South Africa,
{\tt http://sun.ac.za/stias}.}$^{,}$\footnote{On sabbatical leave from the Center for Particle Physics and Phenomenology (CP3),
Institut de Physique Nucl\'eaire, Universit\'e catholique de Louvain (U.C.L.),
2, Chemin du Cyclotron, B-1348 Louvain-la-Neuve, Belgium,
E-mail: {\tt Jan.Govaerts@fynu.ucl.ac.be}.}
and Frederik G. Scholtz$^{\dagger,}$\footnote{E-mail: {\tt FGS@sun.ac.za}.}

\vspace{0.5cm}

$^\dagger${\em Institute of Theoretical Physics}\\
{\em Department of Physics, University of Stellenbosch}\\
{\em Stellenbosch 7600, Republic of South Africa}\\

\vspace{0.7cm}

$^\ddagger${\em UNESCO International Chair in Mathematical Physics and 
Applications (ICMPA)}\\
{\em University of Abomey-Calavi}\\
{\em 072 B.P. 50, Cotonou, Republic of Benin}\\

\vspace{0.7cm}

\begin{abstract}
\noindent
In order to assess possible observable effects of noncommutativity in deformations of quantum
mechanics, all irreducible representations of the noncommutative Heisenberg algebra and Weyl-Heisenberg
group on the two-torus are constructed. This analysis extends the well known situation for the noncommutative torus
based on the algebra of the noncommuting position operators only. When considering the dynamics of a free particle for any
of the identified representations, no observable effect of noncommutativity is implied.

\end{abstract}

\vspace{10pt}

\end{centering} 

\vspace{125pt}

\end{titlepage}

\setcounter{footnote}{0}

\section{Introduction}
\label{Sec1}

\subsection{Motivation}
\label{Sec1.1}

The idea that space and spacetime coordinates may in fact be noncommutative goes as far back as the early
days of quantum mechanics~\cite{Snyder}. In recent years however, it has witnessed greatly renewed interest since
the issue has arisen again within attempts aiming towards a theory for quantum gravity,
whether in the M-theory or Loop Quantum Gravity contexts or more generally deformations of quantum mechanics
at the smallest distance scales. Quantum field theory on noncommutative spacetimes has now grown into a research field
of its own (see e.g. reference~\cite{Rivasseau} and references therein). In the simpler context of mechanical systems, so-called noncommutative quantum mechanics considers
deformations of the ordinary Heisenberg algebra of hermitian operators, $\hat{x}^i$ and $\hat{p}_i$ ($i=1,2,\cdots,d)$,
with for instance in the simplest case a nonvanishing constant space-space commutator,
\begin{equation}
\left[\hat{x}^i,\hat{x}^j\right]=i\theta^{ij}\mathbb{I},\qquad
\left[\hat{x}^i,\hat{p}_j\right]=i\hbar\delta^i_j\mathbb{I},\qquad
\left[\hat{p}_i,\hat{p}_j\right]=0,\qquad i,j=1,2,\cdots,
\label{eq:1}
\end{equation}
the antisymmetric constants $\theta^{ij}=-\theta^{ji}$ thus parametrising such
deformations\footnote{The momentum-momentum commutator may be deformed in a likewise manner
but an appropriate change of variables brings the algebra into the form  (\ref{eq:1}), except
for one singular choice of deformation parameters which shall not be addressed here.}.

It certainly is a legitimate question to identify possible observable consequences of such noncommutative deformations of
quantum mechanics, with deviations from the ordinary situation expected to become apparent at the distance scales
set by the parameters $\theta^{ij}$.  However, when the operators $\hat{x}^i$ and $\hat{p}_i$ are thought of
as cartesian coordinates spanning an Euclidean phase space, the representation theory of the noncommutative Heisenberg
(NC-H) algebra (\ref{eq:1}) is not different from that of the ordinary Heisenberg algebra with
$\theta^{ij}=0$ for which, according to the Stone--von Neumann theorem, there exists a unique representation (up
to unitary transformations). Indeed, by an appropriate linear change of basis in $\hat{x}^i$, the matrix
$\theta^{ij}$ may be $2\times 2$-block diagonalised. Restricted to any such two-dimensional subspace now with
$i,j=1,2$, the NC-H algebra reduces to
\begin{equation}
\left[\hat{x}^i,\hat{x}^j\right]=i\theta\,\epsilon^{ij}\mathbb{I},\qquad
\left[\hat{x}^i,\hat{p}_j\right]=i\hbar\,\delta^i_j\mathbb{I},\qquad
\left[\hat{p}_i,\hat{p}_j\right]=0,\qquad i,j=1,2,
\label{eq:2}
\end{equation}
where, without loss of generality, one assumes $\theta>0$ while $\epsilon^{ij}=\epsilon_{ij}$ is the antisymmetric
symbol with $\epsilon^{12}=+1=\epsilon_{12}$. Considering then the operators defined by the following
linear combinations, corresponding to a Darboux transformation, which brings the commutation relations into canonical form,
\begin{equation}
\hat{X}^i=\hat{x}^i+\frac{\theta}{2\hbar}\epsilon^{ij}\hat{p}_j,
\end{equation}
one recovers the ordinary Heisenberg algebra
\begin{equation}
\left[\hat{X}^i,\hat{X}^j\right]=0,\qquad
\left[\hat{X}^i,\hat{p}_j\right]=i\hbar\delta^i_j\mathbb{I},\qquad
\left[\hat{p}_i,\hat{p}_j\right]=0.
\end{equation}
Since the abstract representation space of the algebra $(\hat{X}^i,\hat{p}_i)$ is unique and coincides in this
construction with that of the original algebra $(\hat{x}^i,\hat{p}_i)$, indeed the quantum states of the deformed NC-H algebra
(\ref{eq:2}) do not differ from those of the ordinary Heisenberg algebra. In other words at the level solely of
the ``kinematics" in an Euclidean configuration space, there are no observable differences between the
commutative, $\theta=0$, and noncommutative, $\theta\ne 0$, versions of the quantum commutation relations.
A similar conclusion holds in the context of quantum field theory on noncommutative spacetime~\cite{Wess}.

One may possibly object to the above argument on the grounds that the plane wave representation of the Heisenberg
algebra does not define a genuine Hilbert space in a strict sense. Consequently the linear transformation between
operator representations could possibly suffer ambiguities related to the behaviour of states at infinity
in the Euclidean plane. However, the restriction to states of Schwartz class is best achieved by considering the
Fock algebra generators
\begin{equation}
b=\frac{1}{\sqrt{2\theta}}\left[\hat{x}^1+i\hat{x}^2\right],\quad
b^\dagger=\frac{1}{\sqrt{2\theta}}\left[\hat{x}^1-i\hat{x}^2\right],\quad
a=b^\dagger+\frac{i}{\hbar}\sqrt{\frac{\theta}{2}}\hat{p}_-,\quad
a^\dagger=b-\frac{i}{\hbar}\sqrt{\frac{\theta}{2}}\hat{p}_+,
\end{equation}
where $\hat{p}_\pm=\hat{p}_1\pm i\hat{p}_2$, such that the only nonvanishing commutators are
\begin{equation}
\left[ b,b^\dagger\right]=\mathbb{I},\qquad
\left[a,a^\dagger\right]=\mathbb{I}.
\end{equation}
Working then in the Hilbert space obtained as the closure of the separable complex vector space spanned by
the Fock (and the coherent) states built out of these two commuting Fock algebras, one obtains wave function
representations of Schwartz class of the NC-H algebra (\ref{eq:2}). It is straightforward to establish that
these representations are isomorphic to the unique ordinary representation of the commutative Heisenberg
algebra with $\theta=0$ by identifying the appropriate changes of bases.

It thus follows that when configuration space is Euclidean any possible observable effect of noncommutativity
must result from the dynamics, namely the specification of a Hamiltonian operator and interactions. However
in the case of a free noncommutative particle with the ordinary nonrelativistic Hamiltonian
\begin{equation}
\hat{H}=\frac{1}{2\mu}\delta^{ij}\hat{p}_i\hat{p}_j,
\label{eq:h1}
\end{equation}
which commutes with the commuting operators $\hat{p}_i$ considered to define the generators of translations in
(the eigenspectrum of) the configuration space coordinate operators $\hat{x}^i$, the energy spectrum, and hence the dynamics itself clearly remains independent
of the noncommutativity parameters $\theta^{ij}$ since the $\hat{p}_i$ eigenspectrum coincides with
that of the commutative Heisenberg algebra. In other words, in the case of an Euclidean configuration space
the manifestation of any observable effects related to noncommutativity is possible at best only in the presence
of interactions (in any case, besides the physical constant $\hbar$, an extra area scale is required
to combine with the noncommutativity parameter $\theta$ to construct physical observables function of $\theta$).
Obviously this is not a welcome feature since it may be difficult to disentangle effects of
interactions from those of noncommutativity. Indeed, such effects may even be physically equivalent in an effective
sense. There are known instances in which interactions in a given energy range within the commutative setting
may be given an equivalent description in terms of noncommuting configuration space variables in the absence
of any interactions safe from the coupling to an applied magnetic field~\cite{sch1,sch2}.

As an alternative one may consider configuration spaces of a topology or geometry different from those of
Euclidean space. Confining even the free particle to some potential well in effect introduces interactions through
boundary conditions at the well. In the presence of coordinate noncommutativity the specification of such
boundary conditions, namely associated to a compact space with boundaries, is not straightforward and requires a dedicated
formulation to be addressed elsewhere. Another form of confinement to a finite volume is through compactification of configuration
space, leading to a finite area $A$. One might then expect that physical observables
may acquire correction factors, which are functions of the ratio $\theta/A$, while the leading order will coincide with the commutative case. The simplest choice for such a compactification
is that of a torus topology. The present work addresses the dynamics of the free particle
on the noncommutative two-torus associated to the noncommutative Heisenberg algebra (\ref{eq:2}). We shall proceed
by first constructing all possible representations of the NC-H algebra for such a geometry, and then
consider the possible dynamics of a free particle.

The rationale for the construction of representations of the algebra (\ref{eq:2}) on
the noncommutative two-torus (NC-2T) is as follows. Any such torus of given geometry may be seen as
the quotient of the Euclidean plane by some abelian lattice group.
In terms of the NC-H algebra (\ref{eq:2}) this lattice group is realised as a specific discrete subgroup
of the exponentiated noncommutative Weyl-Heisenberg (NC-WH) group of which the generators are
$\mathbb{I}$, $\hat{x}^i$ and $\hat{p}_i$ ($i=1,2$). Even though the coordinate operators $\hat{x}^i$
do not commute when $\theta\ne 0$, what is required is only that the group composition law for the
lattice subgroup be abelian, namely additive in the lattice vectors. This requirement should entail
a quantised cocycle condition in the noncommutative case. Having thereby constructed the appropriate lattice group
associated to a given NC-2T geometry, it remains to identify within the unique representation space
of the NC-H algebra (\ref{eq:2}) on the plane those states that are left invariant under the action
of the lattice group, as well as those elements of the full NC-WH group generated by (\ref{eq:2})
which commute with the lattice subgroup of the NC-WH group, namely the normaliser of the lattice subgroup
within the NC-WH group. By construction, the elements of the latter normaliser then map invariant states
into one another in a single-valued manner on the NC-2T. In other words, the set of invariant states defines
a closed representation space for the NC-WH subgroup which commutes with the lattice group characterising the
noncommutative two-torus. The set of such possible representations associated to a given torus geometry
then provides the realm from which to choose a realisation of the noncommutative particle's motion.

In the present case the choice of dynamics, namely of Hamiltonian operator, should reflect the free character
of the particle's motion on the noncommutative two-torus. This is best achieved in an invariant manner, by requiring,
as in the ordinary commutative case, that the Hamiltonian commutes with the generators of space translations.
We take this requirement to define what is meant by a free particle, whether in the commutative or the noncommutative context.
Hence the Hamiltonian will be chosen to be quadratic in the operators which commute with the translation generators.
Since the lattice group is certainly to be constructed in terms of the translation
operators, the action of such a Hamiltonian operator preserves the invariant character of quantum states, hence it
acts within any of the possible representations of the NC-WH group on the NC-2T.

\subsection{Methodology}
\label{Sec1.2}

The construction thus relies entirely, on the one hand, on the choice of lattice vectors specifying the
geometry of the two-torus, and on the other hand, on the specification of the translation operators.
The lattice vectors are to be denoted $e^i_a$ ($a=1,2; i=1,2$) with the following identifications
in the spectrum of $\hat{x}^i$ eigenvalues defining the two-torus\footnote{See the Appendix for a compendium
of useful properties of these lattice vectors and their dual vectors $\tilde{e}^a_i$.},
\begin{equation}
x^i\sim x^i+n^a\,e^i_a,\qquad n^a\in\mathbb{Z}.
\end{equation}
Denoting by $\hat{T}_i$ the translation generators in configuration space, lattice group elements
must be of the form
\begin{equation}
U(n^a)=C(n^a)\,e^{-\frac{i}{\hbar}n^a e_a^i \hat{T}_i},
\label{eq:U}
\end{equation}
where $C(n^a)$ are cocycle factors to be chosen such that the abelian group composition law of the lattice,
additive in the lattice vectors $n^a e_a^i$ and $\ell^a e_a^i$, be obeyed
\begin{equation}
U(n^a)\,U(\ell^a)=U(n^a+\ell^a),\qquad n^a,\ell^a\in\mathbb{Z},
\label{eq:abelianlaw}
\end{equation}
irrespective of whether the operators $\hat{T}_i$ commute with one another or not. The choice of translation
operators $\hat{T}_i$ must be such that their adjoint action on the coordinate operators $\hat{x}^i$ induces
the appropriate lattice shift,
\begin{equation}
U^\dagger(n^a)\,\hat{x}^i\,U(n^a)=\hat{x}^i\,+\,n^a e_a^i\mathbb{I},
\label{eq:torus}
\end{equation}
a condition which requires the property
\begin{equation}
\left[\hat{x}^i,\hat{T}_j\right]=i\hbar\,\delta^i_j\mathbb{I}.
\label{eq:cond1}
\end{equation}
In the ordinary commutative context, the translation generators are taken to coincide with the
conjugate momentum operators, $\hat{T}_i=\hat{p}_i$, in which case these operators commute and
are left invariant by the lattice group spanned by $U(n^a)$. However, in the present context there
is {\it a priori\/} nothing to prevent us from considering more general linear combinations of
the basic operators $\hat{x}^i$ and $\hat{p}_i$ such that the conditions (\ref{eq:cond1})
are met. In the noncommutative case the coordinate operators $\hat{x}^i$ certainly also effect
translations in configuration space, while the commuting momentum operators $\hat{p}_i$ may in fact
then result from linear combinations of $\hat{x}^i$ with originally noncommuting momentum operators.
Certainly in the presence of noncommutativity the distinction between the configuration and
momentum spaces is less clear-cut than in the commutative case, and while one translates in configuration space
translations in momentum space may also be induced on a scale set by $\hbar/\sqrt{\theta}$. From this point of view
we take here the definition of the torus geometry to be given by the relation (\ref{eq:torus})
irrespective of the transformation properties of the momentum operators under the lattice
group operators $U(n^a)$. Note that such a characterisation of the lattice group and the torus
geometry allows even in the commutative case a more general choice for translation operators than
simply the momenta $\hat{p}_i$ as is usually done. Since the possibility offers itself, it
certainly is worth exploring its consequences and possible physical relevance.

Once a choice of translation generators $\hat{T}_i$ has been made in accordance with (\ref{eq:cond1}),
as well as lattice group elements $U(n^a)$ in (\ref{eq:U}) with cocyle factors $C(n^a)$ in compliance with
the abelian group composition law (\ref{eq:abelianlaw}), it is possible to identify the subspace of quantum states
of the unique representation space for the NC-H algebra (\ref{eq:2}) on the noncommutative plane which are invariant
under the lattice group, namely, it is the quotient of the original representation space by the lattice group spanned by $U(n^a)$.
This invariant subspace may also be determined by considering the (non-normalisable) projector (density)
\begin{equation}
\mathbb{P}=\sum_{n^a\in\mathbb{Z}}\,U(n^a)
\label{eq:projector}
\end{equation}
applied on the original representation space.

What then remains to be done is to identify the subgroup of the NC-WH group, generated by the NC-H algebra
(\ref{eq:2}), for which the action on these states closes in a manner consistent with the lattice
group action. More specifically, the general unitary operators representing elements of the NC-WH group generated
by (\ref{eq:2}) are parametrised according to,
\begin{equation}
W(x^i,p_i;\varphi)=\exp\left[i\varphi\mathbb{I}+\frac{i}{\hbar}p_i\hat{X}^i-\frac{i}{\hbar}X^i\hat{p}_i\right]=
\exp\left[i\varphi\mathbb{I}+\frac{i}{\hbar}p_i\hat{x}^i-\frac{i}{\hbar}\left(x^i+\frac{\theta}{\hbar}\epsilon^{ij}p_j\right)
\hat{p}_i\right].
\label{eq:generalW}
\end{equation}
Here
\begin{equation}
X^i=x^i+\frac{\theta}{2\hbar}\epsilon^{ij} p_j,
\end{equation}
with $x^i$, $p_i$ and $\varphi$ (defined modulo $2\pi$) real parameters spanning the NC-WH group.
The reason for this specific choice of parametrisation in terms of the commuting Heisenberg algebra,
associated to $(\hat{X}^i,\hat{p}_i,\mathbb{I})$, is that the adjoint action of the unitary operators
$W(x^i,p_i;\varphi)$ (with $W^\dagger(x^i,p_i;\varphi)=W^{-1}(x^i,p_i;\varphi)=W(-x^i,-p_i;-\varphi)$),
is then indeed such that the operators $\hat{x}^i$ and $\hat{p}_i$ are shifted
by the constant parameters $x^i$ and $p_i$, respectively, and subsequently also their
eigenspectra\footnote{The remaining generator $\mathbb{I}$ of the NC-H
algebra is of course invariant under this adjoint action.},
\begin{equation}
W^\dagger(x^i,p_i;\varphi)\,\hat{x}^i\,W(x^i,p_i;\varphi)=\hat{x}^i\,+\,x^i\mathbb{I},\qquad
W^\dagger(x^i,p_i;\varphi)\,\hat{p}_i\,W(x^i,p_i;\varphi)=\hat{p}_i\,+\,p_i\mathbb{I}.
\end{equation}
The lattice group elements $U(n^a)$ are a particular subclass of these operators with
parameters $(x^i,p_i;\varphi)$ given by specific functions of $n^a\in\mathbb{Z}$.
We thus have
\begin{eqnarray}
U^\dagger(n^a)\,\hat{x}^i\,U(n^a)&=&\hat{x}^i+\Delta_n x^i\mathbb{I},\qquad
\Delta_n x^i=n^a\,e_a^i,\nonumber \\
U^\dagger(n^a)\,\hat{p}_i\,U(n^a)&=&\hat{p}_i\,+\,\Delta_n p_i\mathbb{I},\qquad
\Delta_n p_i=n^a\,\Delta_a p_i,
\end{eqnarray}
where $\Delta_a p_i$ depend on the specific choice of translation generators $\hat{T}_i$.

Requiring now consistency between the action of the NC-WH group elements $W(x^i,p_i;\varphi)$ and the
lattice group elements $U(n^a)$ will restrict the ranges for the NC-WH group parameters $(x^i,p_i;\varphi)$
in such a way that the associated subclass still closes into a subgroup of the original NC-WH group,
namely the noncommutative two-torus Weyl-Heisenberg (NC-2T-WH) group, and commutes with the
lattice group. The action of the NC-2T-WH group then closes on the subspace of invariant states. The latter
condition corresponds to the requirement that, for all $n^a\in\mathbb{Z}$,
\begin{equation}
U(n^a)\,W(x^i,p_i;\varphi)= W(x^i,p_i;\varphi)\,U(n^a),
\label{eq:UW1}
\end{equation}
leading to restrictions on the NC-WH group parameters $(x^i,p_i;\varphi)$.

Furthermore, any such restricted NC-WH group element $W(x^i,p_i;\varphi)$ acting on
an invariant state produces another invariant state which must be single-valued
in lattice shifts of the parameters $(x^i,p_i)$.  Due to the possible nontrivial
cocycle factor $C(n^a)$ in $U(n^a)$, as well as other phase factors arising from
combining the product $U(n^a)W(x^i,p_i;\varphi)$ into a new element of the form
$W(x^i+\Delta_n x^i,p_i+\Delta_n p_i;\varphi')$, this condition
of single-valuedness requires a specific dependence $\varphi(x^i,p_i)$ for the
phase parameter $\varphi$ such that one meets a second restriction
of the form
\begin{equation}
U(n^a)\,W(x^i,p_i;\varphi(x^i,p_i))=W(x^i(n),p_i(n);\varphi(x^i(n),p_i(n)))=
W(x^i,p_i;\varphi)\,U(n^a),
\label{eq:UW2}
\end{equation}
for all $n^a\in\mathbb{Z}$.  Here $x^i(n)=x^i+\Delta_n x^i$ and $p_i(n)=p_i+\Delta_n p_i$.

Provided the two conditions (\ref{eq:UW1}) and (\ref{eq:UW2}) are met,
any invariant state, $U(n^a)|\psi\rangle=|\psi\rangle$, is then mapped into an invariant state,
\begin{equation}
U(n^a)\,W(x^i,p_i;\varphi)|\psi\rangle=W(x^i,p_i;\varphi)\,U(n^a)|\psi\rangle
=W(x^i,p_i;\varphi)|\psi\rangle,
\end{equation}
while any of its NC-2T-WH images is single-valued in any lattice shift of the
group parameters,
\begin{equation}
W(x^i(n),p_i(n);\varphi(x^i(n),p_i(n)))\,|\psi\rangle=
W(x^i,p_i;\varphi(x^i,p_i))\,U(n^a)\,|\psi\rangle=W(x^i,p_i;\varphi(x^i,p_i))\,|\psi\rangle.
\end{equation}

Note that in actual fact none of the above considerations requires the specification
of an inner product on the representation space of the NC-H algebra (\ref{eq:2}) on the
noncommutative plane. It is true that such a structure is required to ensure the hermiticity and unitarity
properties mentioned throughout the above discussion, but, as a matter of fact, one is free to
introduce a different, or new inner product on the final representation space obtained
as the quotient by the lattice group, and still fulfill the
necessary properties of hermiticity and unitarity. This freedom in a (re)definition of the
inner product often allows for normalisable invariant states when the invariant representation space
is discrete or even of finite dimension, in contradistinction to the situation in the
original representation space.

The above general description outlines the approach which is to be developed hereafter. For the purpose of illustration and later comparison with the
noncommutative situation, these considerations are applied in the next Section to the general $d$-dimensional torus in the case of the ordinary commuting Heisenberg
algebra (with $\theta^{ij}=0$ in (\ref{eq:1})).  In Sec.~\ref{Sec3} the same considerations are applied to the ordinary noncommutative
configuration space subalgebra
\begin{equation}
\left[\hat{x}^i,\hat{x}^j\right]=i\theta\,\epsilon^{ij}\mathbb{I},\qquad \theta>0,\qquad
i,j=1,2,
\label{eq:3}
\end{equation}
which does not yet include the momentum operators $\hat{p}_i$. The representation theory of this
structure on the noncommutative two-torus is of course well known~\cite{Connes}. It is rederived here for the purpose
of establishing the consistency of the above construction, and more importantly to show how, by extending
the algebra to include the commuting momentum operators $\hat{p}_i$, the representation theory on the
two-torus becomes drastically different. Section~\ref{Sec4} finally addresses the situation of interest
associated to the algebra (\ref{eq:2}), and establishes the quantised cocycle condition in terms of a integer
quantity $k_0\in\mathbb{Z}$. The latter quantisation condition possesses two distinguished solutions
associated to $k_0=0$, considered in Sec.~\ref{Sec5}, and a generic branch associated to $k_0\ne 0$,
discussed in Sec.~\ref{Sec6}. The results detailed in these three Sections thus provide the representation theory of the noncommutative
two-torus Weyl-Heisenberg group. Finally, Sec.~\ref{Sec7} identifies the free Hamiltonian based on
the considerations mentioned previously, and determines the energy spectrum of the free noncommutative
particle on the two-torus for each of the established representations. The discussion ends
with some Conclusions. An Appendix collects conventions and properties for the two-torus geometry.

\section{The Ordinary General Torus}
\label{Sec2}

In the case of the ordinary commutative Heisenberg algebra on the Euclidean $d$ dimensional plane,
the unitary Weyl-Heisenberg group elements are parametrised according to
\begin{equation}
W(x^i,p_i;\varphi)=\exp\left[i\varphi\mathbb{I}+\frac{i}{\hbar}p_i\hat{x}^i
-\frac{i}{\hbar}x^i\hat{p}_i\right],
\end{equation}
where $x^i,p_i\in\mathbb{R}$ and $\varphi\in[0,2\pi[$ (mod $2\pi$). The group composition law
is\footnote{The identities $e^Ae^B=e^{A+B+[A,B]/2}$ and $e^A B e^{-A}=A+[A,B]$ valid when both $A$ and $B$ commute with their
commutator $[A,B]$, are used throughout.},
\begin{equation}
W(x^i_2,p_{2i};\varphi_2)\,W(x^i_1,p_{1i};\varphi_1)=
e^{\frac{i}{2\hbar}\left(p_{2i}x^i_1-x^i_2p_{1i}\right)}\,
W(x^i_2+x^i_1,p_{2i}+p_{1i};\varphi_2+\varphi_1),
\label{eq:comp1.1}
\end{equation}
from which the following cocycle property follows:
\begin{equation}
W(x^i_1,p_{1i};\varphi_1)\,W(x^i_2,p_{2i};\varphi_2)=
e^{\frac{i}{\hbar}\left(p_{1i}x^i_2-p_{2i}x^i_1\right)}\,
W(x^i_2,p_{2i};\varphi_2)\,W(x^i_1,p_{1i};\varphi_1).
\label{eq:comp1.2}
\end{equation}

This algebra and group are represented in the usual way with as bases, say, the position,
$|x^i\rangle$, or momentum, $|p_i\rangle$, eigenbases of the position, $\hat{x}^i$, and
momentum, $\hat{p}_i$, operators, respectively,
\begin{equation}
\hat{x}^i\,|x^i\rangle=x^i\,|x^i\rangle,\qquad
\hat{p}_i\,|p_i\rangle=p_i\,|p_i\rangle,\qquad
x^i,p_i\in\mathbb{R}.
\end{equation}
Even though the inner product of these bases vectors need not be specified at this stage,
their relative phases may be fixed as follows,
\begin{equation}
|x^i\rangle=e^{-\frac{i}{\hbar}x^i\hat{p}_i}\,|x^i=0\rangle,\qquad
|p_i\rangle=e^{\frac{i}{\hbar}p_i\hat{x}^i}\,|p_i=0\rangle,
\end{equation}
with the properties
\begin{equation}
e^{-\frac{i}{\hbar}x^i_0\hat{p}_i}\,|x^i\rangle=|x^i+x^i_0\rangle,\qquad
e^{\frac{i}{\hbar}p_{0i}\hat{x}^i}\,|p_i\rangle=|p_i+p_{0i}\rangle.
\end{equation}

As translation operators, in the present context, we make the usual choice $\hat{T}_i=\hat{p}_i$,
which is a commuting set of operators. It thus proves convenient henceforth to work in the
momentum eigenbasis $|p_i\rangle$.

The $d$ dimensional torus geometry, $T_d$, is characterised by lattice vectors $e^i_a$ ($a,i=1,2,\cdots,d$),
with their dual vectors $\tilde{e}^a_i$ such that $e_a^i\,\tilde{e}_i^b=\delta^b_a$ and
$\tilde{e}_i^a\,e_a^j=\delta^j_i$, leading to the lattice identification $x^i\sim x^i+n^a e_a^i$
($n^a\in\mathbb{Z}$) defining the torus.  Consequently the lattice group consists of the following elements, providing the
general solution to the composition rule (\ref{eq:abelianlaw}),
\begin{equation}
U(n^a)=e^{2i\pi n^a\lambda_a}\,e^{-\frac{i}{\hbar}n^a e_a^i\hat{p}_i}=
e^{-\frac{i}{\hbar}n^a e_a^i\left(\hat{p}_i-2\pi\hbar\tilde{e}_i^a\lambda_a\right)}=
W(n^a e_a^i,0;2\pi n^a\lambda_a),
\end{equation}
where $\lambda_a\in\mathbb{R}$, defined modulo the integers, are U(1) holonomy factors
labelling inequivalent representations of the Heisenberg algebra on the $T_d$ torus (see e.g. reference ~\cite{Gov1} and references therein),
thus also characterising the cocycle factors $C(n^a)$, $C(n^a)=\exp(2i\pi n^a\lambda_a)$.
Note that lattice shift transformations of the Weyl-Heisenberg group parameters $(x^i,p_i;\varphi)$
are then
\begin{equation}
\Delta_n x^i=n^a e_a^i,\qquad \Delta_n p_i=0.
\end{equation}
It is also obvious that the subspace of invariant states is spanned by all the momentum
eigenstates belonging to the following discrete set
\begin{equation}
|\overline{m}_a\rangle\equiv|\overline{p}_i\rangle,\qquad
\overline{p}_i=2\pi\hbar\tilde{e}_i^a\left[\overline{m}_a+\lambda_a\right],\qquad
\overline{m}_a\in\mathbb{Z}.
\end{equation}
The same identification follows from considering the projection operator (\ref{eq:projector}).

In order to determine the subgroup of Weyl-Heisenberg elements $W(x^i,p_i;\varphi)$ which commutes
with the lattice group, the composition rule (\ref{eq:comp1.1}) implies that
the condition (\ref{eq:UW1}) imposes the restriction
\begin{equation}
W(x^i,p_i;\varphi):\qquad p_i=2\pi\hbar\tilde{e}_i^a m_a,\qquad m_a\in\mathbb{Z}.
\end{equation}
Furthermore, using now (\ref{eq:comp1.2}), the second condition (\ref{eq:UW2}) is obeyed provided
the phase parameter $\varphi$ is restricted to the form,
\begin{equation}
W(x^i,p_i;\varphi):\qquad p_i=2\pi\hbar\tilde{e}_i^a m_a,\qquad
\varphi=\pi x^i\tilde{e}_i^a\left(m_a+2\lambda_a\right).
\end{equation}
Consequently, the Weyl-Heisenberg group for this torus geometry consists of all operators
of the form
\begin{equation}
W_0(x^i,m_a)=W\left(x^i,2\pi\hbar\tilde{e}_i^a m_a;\pi x^i\tilde{e}_i^a(m_a+2\lambda_a)\right)=
e^{2i\pi\tilde{e}_i^a m_a\hat{x}^i}\,
e^{-\frac{i}{\hbar}x^i\left(\hat{p}_i-2\pi\hbar\tilde{e}_i^a\lambda_a\right)},
\end{equation}
labelled by the parameters $x^i\in\mathbb{R}$ and $m_a\in\mathbb{Z}$.
Under lattice shifts, these parameters vary according to
\begin{equation}
\Delta_n x^i=n^a e_a^i,\qquad \Delta_n m_a=0.
\end{equation}
Given the previously specified phase convention for momentum eigenstates,
the representation of the Weyl-Heisenberg group on the space of invariant states is given by
\begin{equation}
W_0(x^i,m_a)|\overline{m}_a\rangle=e^{-2i\pi x^i\tilde{e}_i^a\overline{m}_a}\,
|\overline{m}_a+m_a\rangle.
\end{equation}
Since this action is single-valued under lattice shifts $(\Delta_n x^i=n^a e_a^i,\Delta_n m_a=0)$
of the parameters $(x^i,m_a)$, it suffices to restrict $x^i$ to the fundamental domain of
the lattice defining the torus, $x^i=u^a e_a^i$, $u^a\in[0,1[$. However, all values
$m_a\in\mathbb{Z}$ are required, so that the representation space spanned by
all states $|\overline{m}_a\rangle$ with $\overline{m}_a\in\mathbb{Z}$ is indeed
irreducible under the action of the torus Weyl-Heisenberg group.

Finally, the composition rule of this commutative torus Weyl-Heisenberg group is
\begin{equation}
W_0(x^i_2,m_{2a})\,W_0(x^i_1,m_{1a})=e^{-2i\pi x^i_2\tilde{e}_i^a m_{1a}}\,
W_0(x^i_2+x^i_1,m_{2a}+m_{1a}),
\label{eq:comp1.3}
\end{equation}
from which follows the cocycle property
\begin{equation}
W_0(x^i_1,m_{1a})\,W_0(x^i_2,m_{2a})=
e^{2i\pi\left(x^i_2\tilde{e}_i^am_{1a}-x^i_1\tilde{e}_i^a m_{2a}\right)}\,
W_0(x^i_2,m_{2a})\,W_0(x^i_1,m_{1a}).
\label{comp1.4}
\end{equation}

Hence, for each choice of U(1) holonomy parameters $\lambda_a\in[0,1[$ (mod $\mathbb{Z}$), one obtains
an irreducible countable infinite dimensional representation of the Weyl-Heisenberg group on the $d$ dimensional
torus, spanned by the states $|\overline{m}_a\rangle$, $\overline{m}_a\in\mathbb{Z}$.
One may now (re)specify the inner product on that representation space, ensuring all the required
hermiticity and unitarity properties of operators, with the orthonormalised choice
\begin{equation}
\langle\overline{m}_a|\overline{\ell}_a\rangle=\delta^{(d)}_{\overline{m},\overline{\ell}}.
\end{equation}
That different choices of holonomy parameters $\lambda_a\in[0,1[$ correspond to unitarily inequivalent
representations may be seen, for instance, by noting that the momentum spectrum of invariant states
is given as $\overline{p}_i=2\pi\hbar\tilde{e}_i^a(\overline{m}_a+\lambda_a)$, $\overline{m}_a\in\mathbb{Z}$.
All these results are well known. However, the above discussion serves the purpose of illustrating in a simple case
the general methodology of this paper, while also sharing quite many aspects with parts of the
analysis hereafter.

As a final remark, note that the composition rule (\ref{eq:comp1.3}) allows one to also
readily identify finite or infinite discrete subgroups of the torus Weyl-Heisenberg group
in terms of subsets of the parameters $(x^i,m_a)$ which are closed under the addition
rule defined by (\ref{eq:comp1.3}). The representation space spanned by $|\overline{m}_a\rangle$
may or may not become reducible under such group reductions. However, it is important to keep in mind
that one is then no longer dealing with the torus Weyl-Heisenberg group, but only a subgroup of it,
and possibly then even only a subalgebra of the original Heisenberg algebra spanned
by $\hat{x}^i$, $\hat{p}_i$ and $\mathbb{I}$, as the case may be.

\section{The Ordinary Noncommutative Torus}
\label{Sec3}

Let us now turn to the noncommutative algebra (\ref{eq:3}) spanned only by the three operators
$\hat{x}^i$ ($i=1,2$) and $\mathbb{I}$. Given the two-torus geometry to be considered hereafter, characterised by lattice vectors\footnote{Further properties and conventions are specified
in the Appendix.} $e_a^i$ ($a,i=1,2$), it is convenient to work with the ``rectified" coordinate
operators
\begin{equation}
\hat{u}^a=\hat{x}^i\,\tilde{e}_i^a,\qquad \hat{x}^i=\hat{u}^a\,e_a^i,
\end{equation}
such that\footnote{In the present discussion the ratio $\theta/A$ thus plays a r\^ole
akin to that of Planck's constant $\hbar$ in the one dimensional Heisenberg algebra
$[\hat{x},\hat{p}]=i\hbar$ given the associations $\hat{u}^1\leftrightarrow\hat{x}$
and $\hat{u}^2\leftrightarrow\hat{p}$.}
\begin{equation}
\left[\hat{u}^a,\hat{u}^b\right]=i\frac{\theta}{A}\epsilon^{ab}\,\mathbb{I}.
\end{equation}

The elements of the nonabelian group associated with this noncommutative algebra are
parameterized as follows
\begin{equation}
W(u^a;\varphi)=e^{i\varphi\mathbb{I}-i\frac{A}{\theta}u^a\epsilon_{ab}\hat{u}^b},
\end{equation}
in terms of parameters $u^a\in\mathbb{R}$ and $\varphi\in[0,2\pi[$ (mod $2\pi$) and such that
\begin{equation}
W^\dagger(u^a;\varphi)\,\hat{u}^a\,W(u^a;\varphi)=\hat{u}^a\,+\,u^a\mathbb{I}.
\end{equation}
The group composition law is
\begin{equation}
W(u^a_2;\varphi_2)\,W(u^a_1;\varphi_1)=e^{-\frac{iA}{2\theta}\epsilon_{ab}u^a_2 u^b_1}\,
W(u^a_2+u^a_1;\varphi_2+\varphi_1),
\label{eq:comp2.1}
\end{equation}
from which follows the cocycle property,
\begin{equation}
W(u^a_1;\varphi_1)\,W(u^a_2;\varphi_2)=e^{\frac{iA}{\theta}\epsilon_{ab} u^a_2 u^b_1}\,
W(u^a_2;\varphi_2)\,W(u^a_1;\varphi_1).
\label{eq:comp2.2}
\end{equation}

The representation space of this algebra and group is spanned in terms of either $\hat{u}^1$
or $\hat{u}^2$ eigenstates, $|u^1\rangle_1$ or $|u^2\rangle_2$, respectively,
\begin{equation}
\hat{u}^1\,|u^1\rangle_1=u^1\,|u^1\rangle_1,\qquad
\hat{u}^2\,|u^2\rangle_2=u^2\,|u^2\rangle_2.
\end{equation}
Here again let us only specify the relative phases of these states, but not yet their inner product,
through the definitions
\begin{equation}
|u^1\rangle_1=e^{-\frac{iA}{\theta}u^1\hat{u}^2}\,|u^1=0\rangle_1,\qquad
|u^2\rangle_2=e^{\frac{iA}{\theta}u^2\hat{u}^1}\,|u^2=0\rangle_2,
\end{equation}
a choice which implies the properties
\begin{equation}
e^{-\frac{iA}{\theta}u^1_0\hat{u}^2}\,|u^1\rangle_1=
|u^1+u^1_0\rangle_1,\qquad
e^{\frac{iA}{\theta}u^2_0\hat{u}^1}\,|u^2\rangle_2=
|u^2+u^2_0\rangle_2.
\end{equation}

As translation operators in the present case there is no other choice possible than
$\hat{T}_i=\tilde{e}_i^a\hat{T}_a$ with $\hat{T}_a=\epsilon_{ab}\hat{u}^b$, leading to
the lattice group elements
\begin{equation}
U(n^a)=C(n^a)\,e^{-\frac{iA}{\theta}n^a\epsilon_{ab}\hat{u}^b}.
\end{equation}
The abelian composition law condition (\ref{eq:abelianlaw}) implies the following
cocycle property
\begin{equation}
e^{-\frac{iA}{2\theta}\epsilon_{ab}n^a\ell^b}\,C(n^a)\,C(\ell^a)=C(n^a+\ell^a),
\end{equation}
for which the general solution is given by
\begin{equation}
C(n^a)=e^{-i\pi k_0\,n^1 n^2}\,e^{2i\pi n^a\epsilon_{ab}\lambda^b},
\end{equation}
$k_0\in\mathbb{N}^*$ being a positive natural number in terms of which the
torus area $A$ is quantised in units of $2\pi\theta$,
\begin{equation}
A=2\pi\theta\,k_0,\qquad k_0\in\mathbb{N}^*.
\label{eq:quantisation0}
\end{equation}
This labels a semi-infinite discrete series of representations,
where, once again, $\lambda^a\in[0,1[$ (modulo the integers) are U(1) holonomy parameters
labelling unitarily inequivalent representations of the noncommutative two-torus group
for each value of $k_0$. Given these choices, one thus has
\begin{equation}
U(n^a)=e^{2i\pi k_0 n^2\left(\hat{u}^1-\frac{\lambda^1}{k_0}\right)}\,
e^{-2i\pi k_0 n^1\left(\hat{u}^2-\frac{\lambda^2}{k_0}\right)}=
e^{-2i\pi k_0 n^1\left(\hat{u}^2-\frac{\lambda^2}{k_0}\right)}\,
e^{2i\pi k_0 n^2\left(\hat{u}^1-\frac{\lambda^1}{k_0}\right)}
\end{equation}
with the identification
\begin{equation}
U(n^a)=W(n^a,2\pi n^a\epsilon_{ab}\lambda^b-\pi k_0 n^1 n^2).
\label{eq:identif2}
\end{equation}
Note that under lattice shifts the group parameters $u^a$ transform according to
\begin{equation}
\Delta_n u^a=n^a,\qquad \Delta_a u^b=\delta_a^b.
\end{equation}

Invariant states may be identified in the $|u^1\rangle_1$ or $|u^2\rangle_2$ basis
either by direct construction or by considering the action of the projection operator (\ref{eq:projector}).
In the $|u^2\rangle_2$ basis one finds the following collection of invariant states
\begin{equation}
|\overline{k}^2\rangle\rangle_2=\sum_{\ell^2=-\infty}^{+\infty}\,
e^{-2i\pi\ell^2\lambda^1}\,|\overline{u}^2+\ell^2\rangle_2,\qquad
\overline{u}^2=\frac{\overline{k}^2+\lambda^2}{k_0},\qquad
\overline{k}^2\in\mathbb{Z},
\end{equation}
and likewise in the $|u^1\rangle_1$ basis,
\begin{equation}
|\overline{k}^1\rangle\rangle_1=\sum_{\ell^1=-\infty}^{+\infty}\,
e^{2i\pi\ell^1\lambda^2}\,|\overline{u}^1+\ell^1\rangle_1,\qquad
\overline{u}^1=\frac{\overline{k}^1+\lambda^1}{k_0},\qquad
\overline{k}^1\in\mathbb{Z}.
\end{equation}
However, because of the following properties, for $n^1,n^2\in\mathbb{Z}$,
\begin{equation}
|\overline{k}^2+k_0 n^2\rangle\rangle_2=
e^{2i\pi n^2\lambda^1}\,|\overline{k}^2\rangle\rangle_2,\qquad
|\overline{k}^1+k_0 n^1\rangle\rangle_1=
e^{-2i\pi n^1\lambda^2}\,|\overline{k}^1\rangle\rangle_1,
\label{eq:prop2.1}
\end{equation}
one obtains in each instance a finite $k_0$ dimensional space of
invariant states, labelled by the integers $\overline{k}^2$ or $\overline{k}^1$
defined modulo $k_0$.

Given the identification (\ref{eq:identif2}) and the composition law (\ref{eq:comp2.1}),
it is readily seen that the requirement (\ref{eq:UW1}) is met provided the parameters $u^a$
labelling group transformations are such that
\begin{equation}
u^a=\frac{k^a}{k_0},\qquad k^a\in\mathbb{Z}.
\end{equation}
Under lattice shifts we thus also have 
\begin{equation}
\Delta_n k^a=k_0 n^a,\qquad \Delta_a k^b=k_0 \delta_a^b.
\end{equation}
This equivalence relation for group elements is enforced in a consistent way
by also considering the requirement (\ref{eq:UW2}), which is met provided
the group parameter $\varphi$ is also restricted as follows when $u^a=k^a/k_0$,
\begin{equation}
\varphi(u^a)=\pi\frac{k^1 k^2}{k_0}\,+\,2\pi\epsilon_{ab}\frac{k^a \lambda^b}{k_0}.
\end{equation}
Consequently the noncommutative two-torus group consists of all the operators of the form
\begin{equation}
W_0(k^a)=
W\left(\frac{k^a}{k_0};\pi\frac{k^1 k^2}{k_0}+2\pi\epsilon_{ab}\frac{k^a\lambda^b}{k_0}\right)=
e^{i\pi\frac{k^1 k^2}{k_0}}\,e^{-2i\pi k^a\epsilon_{ab}
\left(\hat{u}^b-\frac{\lambda^b}{k_0}\right)},
\end{equation}
labelled by the integers $k^a\in\mathbb{Z}$. That these integers are defined modulo $k_0$ follows
from the action on the invariant states,
\begin{equation}
W_0(k^a)|\overline{k}^2\rangle\rangle_2=
e^{-2i\pi\frac{k^1\overline{k}^2}{k_0}}\,
e^{-2i\pi\frac{k^2\lambda^1}{k_0}}\,|\overline{k}^2+k^2\rangle\rangle_2,
\end{equation}
\begin{equation}
W_0(k^a)|\overline{k}^1\rangle\rangle_1=
e^{2i\pi\frac{k^1 k^2}{k_0}}\,
e^{2i\pi\frac{k^2\overline{k}^1}{k_0}}\,
e^{2i\pi\frac{k^1\lambda^2}{k_0}}\,|\overline{k}^1+k^1\rangle\rangle_1,
\end{equation}
which are indeed single-valued under lattice shifts $\Delta_n k^a=k_0 n^a$,
provided the properties (\ref{eq:prop2.1}) are taken into account.

The group composition law is
\begin{equation}
W_0(k^a)\,W_0(\ell^a)=e^{-\frac{2i\pi}{k_0}k^1\ell^2}\,W_0(k^a+\ell^a),
\end{equation}
leading to the cocycle property
\begin{equation}
W_0(\ell^a)\,W_0(k^a)=e^{-\frac{2i\pi}{k_0}\epsilon_{ab}\ell^a k^b}\,
W_0(k^a)\,W_0(\ell^a).
\end{equation}

In conclusion, given the quantised torus area $A=2\pi\theta k_0$, the
noncommutative two-torus group is finite dimensional, consists of $k^2_0$
elements, and is generated from the two basic elements $g_1$ and $g_2$ 
given by
\begin{equation}
g_1=W_0(k^1=1,k^2=0),\qquad g_2=W_0(k^1=0,k^2=1),
\end{equation}
which are such that
\begin{equation}
g_2\,g_1=e^{\frac{2i\pi}{k_0}}\,g_1\,g_2.
\end{equation}
The representation space of this group is $k_0$ dimensional, and is spanned by
either the states $|\overline{k}^2\rangle\rangle_2$ or $|\overline{k}^1\rangle\rangle_1$
where $\overline{k}^a=0,1,2,\cdots,k_0-1$ ($a=1,2$). It is possible to define
an inner product on this space, such that all hermiticity and unitarity properties
are obeyed, in terms of the orthonormalisation conditions
\begin{equation}
_2\langle\langle\overline{k}^2|\overline{\ell}^2\rangle\rangle_2=
\delta_{\overline{k}^2,\overline{\ell}^2},\qquad
_1\langle\langle\overline{k}^1|\overline{\ell}^1\rangle\rangle_1=
\delta_{\overline{k}^1,\overline{\ell}^1},
\end{equation}
as well as the overlap functions
\begin{equation}
_1\langle\langle\overline{k}^1|\overline{k}^2\rangle\rangle_2=
\frac{1}{\sqrt{k_0}}\,e^{\frac{2i\pi}{k_0}\left(\overline{k}^1+\lambda^1\right)
\left(\overline{k}^2+\lambda^2\right)}.
\end{equation}

Except for the presence of the U(1) holonomy parameters $\lambda^a\in[0,1[$, these
results are well known~\cite{Connes}. Still they are included here in order to show how they
follow from the methodology outlined in the Introduction, and to contrast them
with the results for the representation theory of the full Weyl-Heisenberg group on the
noncommutative two-torus.

\section{The Noncommutative Weyl-Heisenberg Algebra on the Torus}
\label{Sec4}

Let us now turn to the full noncommutative Heisenberg algebra (\ref{eq:2}) on
the noncommutative Euclidean plane. We define the following
basis of operators in terms of the lattice vectors $e_a^i$ defining
the two-torus geometry to be considered presently,
\begin{equation}
\hat{u}^a=\hat{x}^i\,\tilde{e}_i^a,\qquad
\hat{v}_a=e_a^i\,\hat{p}_i;\qquad
\hat{x}^i=\hat{u}^a\,e_a^i,\qquad
\hat{p}_i=\tilde{e}_i^a\,\hat{v}_a.
\end{equation}
The NC-H algebra then reads
\begin{equation}
\left[\hat{u}^a,\hat{u}^b\right]=i\frac{\theta}{A}\epsilon^{ab}\mathbb{I},\qquad
\left[\hat{u}^a,\hat{v}_b\right]=i\hbar\delta^a_b\mathbb{I},\qquad
\left[\hat{v}_a,\hat{v}_b\right]=0.
\end{equation}
Introducing also
\begin{equation}
\hat{U}^a=\hat{X}^i\tilde{e}_i^a=\hat{u}^a+\frac{\theta}{2A\hbar}\epsilon^{ab}\,\hat{v}_b,\qquad
\hat{u}^a=\hat{U}^a-\frac{\theta}{2A\hbar}\epsilon^{ab}\hat{v}_b,
\end{equation}
the algebra becomes of the ordinary commutative type,
\begin{equation}
\left[\hat{U}^a,\hat{U}^b\right]=0,\qquad
\left[\hat{U}^a,\hat{v}_b\right]=i\hbar\,\delta^a_b\mathbb{I},\qquad
\left[\hat{v}_a,\hat{v}_b\right]=0.
\end{equation}
Hence the unique representation space is spanned either by $\hat{U}^a$ or $\hat{v}_a$ eigenstates
with eigenvalues $U^a\in\mathbb{R}$ or $v_a\in\mathbb{R}$, respectively,
\begin{equation}
\hat{U}^a\,|U^a\rangle=U^a\,|U^a\rangle,\qquad
\hat{v}_a\,|v_a\rangle=v_a\,|v_a\rangle.
\end{equation}
Once again our convention for relative phases is such that
\begin{equation}
|U^a\rangle=e^{-\frac{i}{\hbar}U^a\hat{v}_a}\,|U^a=0\rangle,\qquad
|v_a\rangle=e^{\frac{i}{\hbar}v_a\hat{U}^a}\,|v_a=0\rangle,
\end{equation}
and hence
\begin{equation}
e^{-\frac{i}{\hbar}U^a_0\hat{v}_a}\,|U^a\rangle=|U^a+U^a_0\rangle,\qquad
e^{\frac{i}{\hbar}v_{0a}\hat{U}^a}\,|v_a\rangle=|v_a+v_{0a}\rangle.
\end{equation}

The noncommutative Weyl-Heisenberg group elements are parametrised according to
\begin{equation}
W(U^a,v_a;\varphi)=\exp\left[i\varphi\mathbb{I}+\frac{i}{\hbar}v_a\hat{U}^a-
\frac{i}{\hbar}U^a\hat{v}_a\right]=
\exp\left[i\varphi\mathbb{I}+\frac{i}{\hbar}v_a\hat{u}^a-\frac{i}{\hbar}
\left(u^a+\frac{\theta}{A\hbar}\epsilon^{ab}v_b\right)\hat{v}_a\right],
\end{equation}
where $u^a,U^a,v_a\in\mathbb{R}$ with the relations
\begin{equation}
U^a=u^a+\frac{\theta}{2A\hbar}\epsilon^{ab}\,v_b,\qquad
u^a=U^a-\frac{\theta}{2A\hbar}\epsilon^{ab} v_b.
\end{equation}
These operators are such that
\begin{eqnarray}
W^\dagger(U^a,v_a;\varphi)\,\hat{u}^a\,W(U^a,v_a;\varphi)&=&\hat{u}^a+u^a\mathbb{I},\nonumber \\
W^\dagger(U^a,v_a;\varphi)\,\hat{U}^a\,W(U^a,v_a;\varphi)&=&\hat{U}^a+U^a\mathbb{I}, \\
W^\dagger(U^a,v_a;\varphi)\,\hat{v}_a\,W(U^a,v_a;\varphi)&=&\hat{v}_a+v_a\mathbb{I},\nonumber
\label{eq:shifts3.1}
\end{eqnarray}
while their group composition law is
\begin{equation}
W(U^a_2,v_{2a};\varphi_2)\,W(U^a_1,v_{1a};\varphi_1)=
e^{\frac{i}{2\hbar}\left(v_{2a}U^a_1-U^a_2v_{1a}\right)}\,
W(U^a_2+U^a_1,v_{2a}+v_{1a};\varphi_2+\varphi_1),
\label{eq:comp3.1}
\end{equation}
implying the cocycle property
\begin{equation}
W(U^a_1,v_{1a};\varphi_1)\,W(U^a_2,v_{2a};\varphi_2)=
e^{\frac{i}{\hbar}\left(v_{1a}U^a_2-v_{2a}U^a_1\right)}\,
W(U^a_2,v_{2a};\varphi_2)\,W(U^a_1,v_{1a};\varphi_1).
\label{eq:comp3.2}
\end{equation}

For the reasons mentioned in the Introduction, one may
consider as translation operators $\hat{T}_i$ some arbitrary linear combination of $\hat{p}_i$ and $\epsilon_{ij}\hat{x}^j$, which both effect translations in the coordinate operators $\hat{x}^i$. Specifically, when imposing
also the condition (\ref{eq:cond1}), the choice to be made is
\begin{equation}
\hat{T}_i=\left(1-\frac{\beta\theta}{\hbar}\right)\hat{p}_i\,+\,\beta\epsilon_{ij}\hat{x}^j,
\end{equation}
where $\beta\in\mathbb{R}$ is an arbitrary real variable, with appropriate physical dimension,
parametrising the freedom in the choice of translation operators. Note that even in the commutative
case, $\theta=0$, a nonvanishing $\beta$ deforms the choice of translation group compared
to the usual choice $\hat{T}_i=\hat{p}_i$, corresponding to $\beta=0$.
When $\theta\ne 0$, the value $\beta=\hbar/\theta$ corresponds to a choice
of translation operators which is that of the ordinary noncommutative torus of
Sec.~\ref{Sec3}.

For later analysis, it is convenient to rather use the ``rectified" translation operators
\begin{equation}
\hat{T}_a=e_a^i\,\hat{T}_i=\left(1-\frac{\beta\theta}{\hbar}\right)\hat{v}_a+
\beta A\epsilon_{ab}\,\hat{u}^b=\left(1-\frac{\beta\theta}{2\hbar}\right)\hat{v}_a+
\beta A\epsilon_{ab}\,\hat{U}^b.
\end{equation}
The relevant commutation relations are found to be
\begin{equation}
\left[\hat{u}^a,\hat{T}_b\right]=i\hbar\,\delta^a_b\mathbb{I},\qquad
\left[\hat{U}^a,\hat{T}_b\right]=i\hbar\left(1-\frac{\beta\theta}{2\hbar}\right)\delta^a_b
\mathbb{I},\qquad
\left[\hat{v}_a,\hat{T}_b\right]=i\hbar\,\beta A\epsilon_{ab}\mathbb{I},
\end{equation}
while the algebra of the translation group is
\begin{equation}
\left[\hat{T}_a,\hat{T}_b\right]=i\hbar\,2\beta A\left(1-\frac{\beta\theta}{2\hbar}\right)
\epsilon_{ab}\mathbb{I}.
\end{equation}

In view of the expression for $\hat{T}_a$, it proves useful to also introduce the operators
\begin{equation}
\hat{Q}_a=\left(1-\frac{\beta\theta}{2\hbar}\right)\hat{v}_a-\beta A\epsilon_{ab }\,\hat{U}^b=
\hat{v}_a-\beta A\epsilon_{ab}\,\hat{u}^b,
\end{equation}
which are such that
\begin{equation}
\left[\hat{u}^a,\hat{Q}_b\right]=i\hbar\left(1-\frac{\beta\theta}{\hbar}\right)\delta^a_b\mathbb{I},\qquad
\left[\hat{U}^a,\hat{Q}_b\right]=i\hbar\left(1-\frac{\beta\theta}{2\hbar}\right)\delta^a_b\mathbb{I},\qquad
\left[\hat{v}_a,\hat{Q}_b\right]=-i\hbar\,\beta A\epsilon_{ab}\mathbb{I},
\end{equation}
and
\begin{equation}
\left[\hat{Q}_a,\hat{Q}_b\right]=-i\hbar\,2\beta A\left(1-\frac{\beta\theta}{2\hbar}\right)\epsilon_{ab}\mathbb{I}.
\end{equation}
From this follows the important result
\begin{equation}
\left[\hat{T}_a,\hat{Q}_b\right]=0.
\end{equation}
However, since
\begin{equation}
\hat{Q}_a+\hat{T}_a=2\left(1-\frac{\beta\theta}{2\hbar}\right)\hat{v}_a,\qquad
\hat{Q}_a-\hat{T}_a=-2\beta A\epsilon_{ab}\,\hat{U}^b,
\end{equation}
it is only when $2\beta A(1-\beta\theta/(2\hbar))\ne 0$ that the algebra $(\hat{Q}_a,\hat{T}_a,\mathbb{I})$
is equivalent to any of the equivalent algebras $(\hat{x}^i,\hat{p}_i,\mathbb{I})$,
$(\hat{u}^a,\hat{v}_a,\mathbb{I})$ or $(\hat{U}^a,\hat{v}_a,\mathbb{I})$. Under this condition one has the
inverse relations
\begin{equation}
\hat{U}^a=\frac{1}{\beta A}\frac{1}{2}\epsilon^{ab}\left[\hat{Q}_b-\hat{T}_b\right],\qquad
\hat{v}_a=\frac{1}{\left(1-\frac{\beta\theta}{2\hbar}\right)}\frac{1}{2}
\left[\hat{Q}_a+\hat{T}_a\right].
\end{equation}

Finally, under the same condition, $2\beta A(1-\beta\theta/(2\hbar))\ne 0$, the following
expression is also of use when considering the NC-WH group elements introduced previously,
\begin{equation}
v_a\hat{U}^a-U^a\hat{v}_a =\frac{1}{2\beta A\left(1-\frac{\beta\theta}{2\hbar}\right)}
\left[Q_a\,\epsilon^{ab}\hat{Q}_b\,-\,T_a\,\epsilon^{ab}\hat{T}_b\right],
\label{eq:identity2}
\end{equation}
where
\begin{equation}
T_a=\left(1-\frac{\beta\theta}{2\hbar}\right)v_a+\beta A\epsilon_{ab} U^b,
\end{equation}
\begin{equation}
Q_a=\left(1-\frac{\beta\theta}{2\hbar}\right)v_a-\beta A\epsilon_{ab}U^b.
\end{equation}
In addition to the adjoint actions in (\ref{eq:shifts3.1}), one also finds
\begin{eqnarray}
W^\dagger(U^a,v_a;\varphi)\,\hat{T}_a\,W(U^a,v_a;\varphi)&=&\hat{T}_a+T_a\,\mathbb{I},\nonumber \\
W^\dagger(U^a,v_a;\varphi)\,\hat{Q}_a\,W(U^a,v_a;\varphi)&=&\hat{Q}_a+Q_a\,\mathbb{I}.
\end{eqnarray}

Turning to the translation group elements
\begin{equation}
U(n^a)=C(n^a)\,e^{-\frac{i}{\hbar}n^a\hat{T}_a},
\end{equation}
the abelian composition law condition (\ref{eq:abelianlaw}) implies the
cocycle condition
\begin{equation}
e^{-\frac{i}{2\hbar}2\beta A\left(1-\frac{\beta\theta}{2\hbar}\right)\epsilon_{ab}n^a\ell^b}\,
C(n^a)\,C(\ell^a)=C(n^a+\ell^a).
\end{equation}
The general solution is of the form
\begin{equation}
C(n^a)=e^{-i\pi k_0 n^1 n^2}\,e^{2i\pi n^a\epsilon_{ab}\lambda^b},
\end{equation}
where $\lambda^a\in[0,1[$ (modulo the integers) are, once again, U(1) holonomy parameters,
while $k_0\in\mathbb{Z}$ is an integer such that
\begin{equation}
2\beta A\left(1-\frac{\beta\theta}{2\hbar}\right)=2\pi\hbar\,k_0,\qquad k_0\in\mathbb{Z},\qquad \beta\in\mathbb{R}.
\label{eq:quantised}
\end{equation}
This condition generalises the area quantisation condition (\ref{eq:quantisation0}), which applies to the
ordinary noncommutative torus discussed in Sec.~\ref{Sec3}, to the noncommutative Heisenberg algebra
in the presence of the $\beta$ parameter. In particular, for the choice
$\beta=\hbar/\theta$, the integer $k_0$ must again be such that $A=2\pi\theta\,k_0$.

As a function of $A$, $\theta$ and $k_0$, the allowed values for $\beta$ are thus
\begin{equation}
\beta=\frac{\hbar}{\theta}\left[1\pm\sqrt{1-\frac{2\pi\theta}{A}k_0}\right],\qquad
k_0\le\frac{A}{2\pi\theta},\qquad k_0\in\mathbb{Z}.
\end{equation}
The choice $\beta=\hbar/\theta$ corresponds precisely to the degenerate case $A=2\pi\theta\,k_0$
with $k_0>0$. The value $k_0=0$ is associated to the two distinct situations
$\beta=0$ or $\beta=2\hbar/\theta$, namely $2\beta A(1-\beta\theta/(2\hbar))=0$.
This is also the situation when the translation generators $\hat{T}_a$ commute. For any fixed
positive $k_0>0$, as the area $A$ increases continuously from the minimal value
$2\pi\theta\,k_0$, the two above branches of $\beta$ values either decrease or increase
from $\beta=\hbar/\theta$ towards the two singular values $\beta=0$ or $\beta=2\hbar/\theta$,
respectively. Hence the interval $\beta\in]0,2\hbar/\theta[$ is certainly distinguished
when $k_0\ne 0$ for any finite area $A$, while for a finite area $A$ the two end points of that
interval correspond only to the case with $k_0=0$. Strictly negative values of $k_0$ correspond to
$\beta$ values outside the interval $[0,2\hbar/\theta]$. Note that in the commutative case, the
only surviving branch is such that
\begin{equation}
\theta=0:\qquad \beta=\frac{\pi\hbar}{A}\,k_0,\qquad k_0\in\mathbb{Z}.
\end{equation}
Thus, besides the ordinary choice $\beta=0$ corresponding to $k_0=0$, there still exist
many other possibilities for a choice of translation operators. Of course, it is only
when $\beta=0$ that the momentum operators $\hat{p}_i$ are not affected by translations
in configuration space.

In conclusion, the lattice group defining the noncommutative two-torus geometry is generated by the
following elements of the NC-WH group,
\begin{equation}
U(n^a)=e^{-i\pi k_0 n^1 n^2}\,e^{2i\pi n^a\epsilon_{ab}\lambda^b}\,
e^{-\frac{i}{\hbar}n^a\hat{T}_a}=
W\left((1-\frac{\beta\theta}{2\hbar})n^a,\beta A\epsilon_{ab}n^b;
2\pi n^a\epsilon_{ab}\lambda^b-\pi k_0 n^1 n^2\right).
\label{eq:Ugen}
\end{equation}
In particular, the translation shifts induced for each of the operators of
interest, $U^\dagger(n^a)\hat{\cal O} U(n^a)=\hat{\cal O}+\Delta_n{\cal O}\,\mathbb{I}$,
are such that
\begin{eqnarray}
\hat{\cal O}=\hat{u}^a &:&\qquad \Delta_n u^a=n^a,\nonumber \\
\hat{\cal O}=\hat{U}^a &:&\qquad \Delta_n U^a=\left(1-\frac{\beta\theta}{2\hbar}\right)n^a,\nonumber \\
\hat{\cal O}=\hat{v}_a &:&\qquad \Delta_n v_a=\beta A \epsilon_{ab} n^b, \\
\hat{\cal O}=\hat{T}_a &:&\qquad \Delta_n T_a=
2\beta A\left(1-\frac{\beta\theta}{2\hbar}\right)\epsilon_{ab} n^b=2\pi\hbar k_0 \epsilon_{ab} n^b,\nonumber \\
\hat{\cal O}=\hat{Q}_a &:&\qquad \Delta_n Q_a=0. \nonumber
\end{eqnarray}
In order to proceed now with the construction of representations of the NC-2T-WH group,
one needs to consider separately the distinct cases $k_0=0$ from the generic situation
with $k_0\ne 0$.

\section{The Distinct Representations with $k_0=0$}
\label{Sec5}

\subsection{The point $\beta=0$}
\label{Sec5.1}

The degenerate case $\beta=0$ corresponds to the choices
\begin{equation}
\hat{T}_a=\hat{v}_a,\qquad
\hat{Q}_a=\hat{v}_a,\qquad k_0=0.
\end{equation}
The lattice group then consists of the commuting elements
\begin{equation}
U(n^a)=e^{-\frac{i}{\hbar}n^a\left(\hat{v}_a-2\pi\hbar\lambda_a\right)}
=W(n^a,0;2\pi n^a\lambda_a),\qquad
\lambda_a=\epsilon_{ab}\lambda^b.
\end{equation}
Consequently, the situation is comparable to the discussion in Sec.~\ref{Sec2}
for the commuting Weyl-Heisenberg group. In particular, whether by considering the
projection operator (\ref{eq:projector}) or the above expression, it is clear that
the subspace of invariant states is spanned by the following discrete set of $\hat{v}_a$ eigenstates,
\begin{equation}
|\overline{m}_a\rangle\equiv|\overline{v}_a\rangle,\qquad
\overline{v}_a=2\pi\hbar\left(\overline{m}_a+\lambda_a\right),\qquad
\overline{m}_a\in\mathbb{Z}.
\end{equation}

Considering now the NC-WH group elements $W(U^a,v_a;\varphi)$, based on the composition
law (\ref{eq:comp3.1}) it readily follows that the invariance condition (\ref{eq:UW1})
implies the restriction
\begin{equation}
v_a=2\pi\hbar\,m_a,\qquad m_a\in\mathbb{Z}.
\end{equation}
Furthermore, for any such value of $v_a$, the invariance condition (\ref{eq:UW2})
leads to the following choice for the group parameter $\varphi$,
\begin{equation}
\varphi(U^a,m_a)=\pi U^a\left(m_a+2\lambda_a\right).
\end{equation}
Note that under lattice shifts the parameters $(U^a,m_a)$ transform according to
\begin{equation}
\Delta_n U^a=n^a,\qquad \Delta_n m_a=0.
\end{equation}
Consequently, in the case $\beta=0$ the two-torus noncommutative Weyl-Heisenberg group
consists of all the following elements
\begin{equation}
W_0(U^a,m_a)=W\left(U^a,2\pi\hbar m_a;\pi U^a(m_a+2\lambda_a)\right)=
e^{2i\pi m_a\hat{U}^a}\,e^{-\frac{i}{\hbar}U^a(\hat{v}_a-2\pi\hbar\lambda_a)},
\end{equation}
where $U^a\in[0,1[$ (modulo the integers) and $m_a\in\mathbb{Z}$. The representation
of the group on the space of invariant states is
\begin{equation}
W_0(U^a,m_a)|\overline{m}_a\rangle=e^{-2i\pi U^a\overline{m}_a}\,|\overline{m}_a+m_a\rangle,
\end{equation}
which is indeed single-valued under lattice shifts of the group parameters.
Finally, the group composition law is
\begin{equation}
W_0\left(U^a_2,m_{2a}\right)\,W_0\left(U^a_1,m_{1a}\right)=e^{-2i\pi m_{1a} U^a_2}\,
W_0\left(U^a_2+U^a_1,m_{2a}+m_{1a}\right),
\end{equation}
which leads to the cocycle property
\begin{equation}
W_0\left(U^a_1,m_{1a}\right)\,W_0\left(U^a_2,m_{2a}\right)=
e^{2i\pi(U^a_2 m_{1a}-U^a_1 m_{2a})}\,
W_0\left(U^a_2,m_{2a}\right)\,W_0\left(U^a_1,m_{1a}\right).
\end{equation}

In conclusion in the case $\beta=0$, the representation of the noncommutative two-torus Weyl-Heisenberg
algebra is discrete infinite dimensional, and essentially coincides with the representation of the
Weyl-Heisenberg group on the commutative torus discussed in Sec.~\ref{Sec2}.

\subsection{The point $\beta=2\hbar/\theta$}
\label{sec5.2}

The value $\beta=2\hbar/\theta$ corresponds to the second branch with $k_0=0$ and applies
only in the noncommutative case, $\theta\ne 0$. This situation corresponds to the choice
\begin{equation}
\hat{T}_a=\frac{2A\hbar}{\theta}\epsilon_{ab}\hat{U}^b,\qquad
\hat{Q}_a=-\hat{T}_a,
\end{equation}
with the commutative translation algebra
\begin{equation}
\left[\hat{T}_a,\hat{T}_b\right]=0.
\end{equation}
The lattice group thus consists of the commuting elements
\begin{equation}
U(n^a)=e^{-2i\frac{A}{\theta} n^a\epsilon_{ab}(\hat{U}^b-\pi\frac{\theta}{A}\lambda^b)}=
W\left(0,\frac{2A\hbar}{\theta}\epsilon_{ab}n^b;2\pi n^a\epsilon_{ab}\lambda^b\right),
\end{equation}
which induce the following lattice shifts
\begin{equation}
\Delta_n u^a=n^a,\quad
\Delta_n U^a=0,\quad
\Delta_n v_a=\frac{2A\hbar}{\theta}\epsilon_{ab} n^b,\quad
\Delta_n T_a=0,\quad
\Delta_n Q_a=0.
\end{equation}

From the above expression, or by considering the action of the projection operator
(\ref{eq:projector}), invariant states are seen to be spanned by the following discrete
set of $\hat{U}^a$ eigenstates,
\begin{equation}
|\overline{k}^a\rangle\equiv|\overline{U}^a\rangle:\qquad
\overline{U}^a=\frac{\pi\theta}{A}\left(\overline{k}^a+\lambda^a\right),\qquad
\overline{k}^a\in\mathbb{Z}.
\end{equation}

Considering now the invariance condition (\ref{eq:UW1}), based on the composition
law (\ref{eq:comp3.1}), the following restriction arises for the parameters
of the NC-WH group elements $W(U^a,v_a;\varphi)$,
\begin{equation}
U^a=\frac{\pi\theta}{A}\,k^a,\qquad k^a\in\mathbb{Z}.
\end{equation}
Furthermore, given such a value for $U^a$, the requirement (\ref{eq:UW2}) leads
to the following choice of parameter $\varphi$ for those NC-WH transformations,
\begin{equation}
W(U^a,v_a;\varphi):\qquad
\varphi(k^a,v_a)=-\frac{\pi\theta}{2A\hbar}\left(k^a+2\lambda^a\right)\,v_a.
\end{equation}
Note that under lattice shifts the parameters $(k^a,v_a)$ transform according to
\begin{equation}
\Delta_a k^a=0,\qquad \Delta_n v_a=\frac{2A\hbar}{\theta}\epsilon_{ab} n^b.
\end{equation}
Consequently, in the case $\beta=2\hbar/\theta$ the two-torus noncommutative Weyl-Heisenberg
group consists of all the elements
\begin{equation}
W_0(k^a,v_a)=W\left(\frac{\pi\theta}{A}k^a,v_a;-\frac{\pi\theta}{2A\hbar}\left(k^a+2\lambda^a\right)v_a\right)=
e^{-i\frac{\pi\theta}{A\hbar} k^a\hat{v}_a}\,e^{\frac{i}{\hbar}(\hat{U}^a-\frac{\pi\theta}{A}\lambda^a)},
\end{equation}
where $v_a\in[0,2A\hbar/\theta[$ (modulo $2A\hbar/\theta$) and $k^a\in\mathbb{Z}$.
The action of the group on the invariant states is
\begin{equation}
W_0(k^a,v_a)|\overline{k}^a\rangle=e^{i\frac{\pi\theta}{A\hbar}v_a\overline{k}^a}\,
|\overline{k}^a+k^a\rangle,
\end{equation}
which is indeed single-valued in lattice shifts of the group parameters $(k^a,v_a)$.
The group composition law is
\begin{equation}
W_0(k^a_2,v_{2a})\,W_0(k^a_1,v_{1a})=e^{i\frac{\pi\theta}{A\hbar}v_{2a}k^a_1}\,W_0(k^a_2+k^a_1,v_{2a}+v_{1a}),
\end{equation}
from which follows the cocycle property
\begin{equation}
W_0(k^a_1,v_{1a})\,W_0(k^a_2,v_{2a})=e^{i\frac{\pi\theta}{A\hbar}(v_{1a}k^2_2-v_{2a}k^a_1)}\,
W_0(k^a_2,v_{2a})\,W_0(k^a_1,v_{1a}).
\end{equation}

In conclusion, in the case $\beta=2\hbar/\theta$ the noncommutative two-torus Weyl-Heisenberg
group possesses a single discrete infinite dimensional representation, very similar to
the one for $\beta=0$, except that in this case it is in the dual eigenspace of the $\hat{U}^a$ operators.

\section{The Generic Representations with $k_0\ne 0$}
\label{Sec6}

When $k_0\ne 0$ the lattice group elements are given in (\ref{eq:Ugen}). A basis of
invariant states may be constructed in either the $\hat{U}^a$ or $\hat{v}_a$ eigensectors.
In the latter case, let us introduce the notation
\begin{equation}
|\overline{\nu}^a\rangle\equiv|\overline{v}_a\rangle:\qquad
\overline{v}_a=\frac{\beta A}{k_0}\epsilon_{ab}\left(\overline{\nu}^b+\lambda^b\right).
\end{equation}
Considering either the projection operator (\ref{eq:projector}) or the action of
the lattice group on the states $|v_a\rangle$, it is found that invariant states are
spanned by the combinations
\begin{equation}
|\overline{\nu}^a\rangle\rangle=\sum_{\ell^a\in\mathbb{Z}}
e^{i\pi k_0 \ell^1 \ell^2 + i\pi\ell^a\epsilon_{ab}\lambda^b
-i\pi n\ell^a\epsilon_{ab}\overline{\nu}^b}\,|\overline{\nu}^a+k_0\ell^a\rangle,
\end{equation}
which possess, for $n^a\in\mathbb{Z}$, the following property,
\begin{equation}
|\overline{\nu}^a+k_0 n^a\rangle\rangle=
e^{i\pi k_0 n^1 n^2 -i\pi n^a\epsilon_{ab}\lambda^b + i\pi n^a\epsilon_{ab}\overline{\nu}^b}\,
|\overline{\nu}^a\rangle\rangle.
\end{equation}
This shows that the two parameters $\overline{\nu}^a$ are indeed each defined modulo $k_0$.

Likewise in the $\hat{U}^a$ eigensector, let us introduce the notation
\begin{equation}
|\overline{\mu}^a\rangle\equiv|\overline{U}^a\rangle:\qquad
\overline{U}^a=\frac{1}{k_0}\left(1-\frac{\beta\theta}{2\hbar}\right)
\left(\overline{\mu}^a+\lambda^a\right).
\end{equation}
It is then found that invariant states are spanned by the combinations
\begin{equation}
|\overline{\mu}^a\rangle\rangle=\sum_{\ell^a\in\mathbb{Z}}
e^{i\pi k_0\ell^1 \ell^1 + i\pi\ell^a\epsilon_{ab}\lambda^b
-i\pi\ell^a\epsilon_{ab}\overline{\mu}^b}\,|\overline{\mu}^a+k_0\ell^a\rangle,
\end{equation}
which possess, for $n^a\in\mathbb{Z}$, the properties
\begin{equation}
|\overline{\mu}^a+k_0 n^a\rangle\rangle=
e^{i\pi k_0 n^1 n^2 - i\pi n^a\epsilon_{ab}\lambda^b + i\pi n^a\epsilon_{ab}\overline{\mu}^b}\,
|\overline{\mu}^a\rangle\rangle,
\end{equation}
showing that the two parameters $\overline{\mu}^a$ are indeed each defined modulo $k_0$.

Considering the general NC-WH operators $W(U^a,v_a;\varphi)$ and
their group composition law (\ref{eq:comp3.1}), the invariance condition
(\ref{eq:UW1}) imposes the restriction
\begin{equation}
T_a=\left(1-\frac{\beta\theta}{2\hbar}\right)v_a+\beta A\epsilon_{ab}U^b=2\pi\hbar\epsilon_{ab}\,k^b,\qquad
k^a\in\mathbb{Z},
\end{equation}
whereas the linearly independent combination
\begin{equation}
Q_a=\left(1-\frac{\beta\theta}{2\hbar}\right)v_a-\beta A\epsilon_{ab}U^b=2\pi\hbar\epsilon_{ab}\,\rho^a,\qquad
\rho^a\in\mathbb{R},
\end{equation}
is left arbitrary. Note that lattice shifts induce the following transformations for the variables $(k^a,\rho^a)$,
\begin{equation}
\Delta_n k^a=k_0 n^a,\qquad \Delta_n \rho^a=0.
\end{equation}
Furthermore, when this restriction is met, the second invariance condition (\ref{eq:UW2}) leads to the
following choice for the group parameter $\varphi$,
\begin{equation}
\varphi(k^a,\rho^a)=-\frac{\pi}{k_0}k^1 k^2 + \frac{2\pi}{k_0}\epsilon_{ab} k^a \lambda^b.
\end{equation}
Consequently, the NC-WH group elements are given by
\begin{equation}
W_0(k^a,\rho^a)=W\left(U^a,v_a;-\frac{\pi}{k_0}k^1 k^2+\frac{2\pi}{k_0}\epsilon_{ab} k^a\lambda^b\right)
\end{equation}
where
\begin{equation}
U^a=\left(1-\frac{\beta\theta}{2\hbar}\right)\frac{1}{k_0}\left(k^a-\rho^a\right),\qquad
v_a=\frac{\beta A}{k_0}\epsilon_{ab}\left(k^a+\rho^a\right).
\end{equation}
As a matter of fact one also has (see (\ref{eq:identity2}))
\begin{equation}
W_0(k^a,\rho^a)=e^{-i\frac{\pi}{k_0}k^1 k^2+2i\pi\epsilon_{ab}\frac{k^a}{k_0}\lambda^b}\,
e^{\frac{i}{\hbar}\frac{\rho^a}{k_0}\hat{Q}_a}\,e^{-\frac{i}{\hbar}\frac{k^a}{k_0}\hat{T}_a},
\label{eq:Wgen}
\end{equation}
where $k^a\in\mathbb{Z}$ modulo $k_0$ and $\rho^a\in\mathbb{R}$. The representation of the
group is such that when acting on invariant states one finds
\begin{equation}
W_0(k^a,\rho^a)|\overline{\nu}^a\rangle\rangle=
e^{-i\frac{\pi}{k_0}k^1 k^2-i\frac{\pi}{k_0}\epsilon_{ab}k^a(\overline{\nu}^b-\lambda^b)
-i\frac{\pi}{k_0}\epsilon_{ab}(\overline{\nu}^a+k^a+\lambda^a)\rho^b}\,
|\overline{\nu}^a+k^a+\rho^a\rangle\rangle,
\end{equation}
\begin{equation}
W_0(k^a,\rho^a)|\overline{\mu}^a\rangle\rangle=
e^{-i\frac{\pi}{k_0}k^1 k^2-i\frac{\pi}{k_o}\epsilon_{ab} k^a(\overline{\mu}^b-\lambda^b)
+i\frac{\pi}{k_0}\epsilon_{ab}(\overline{\mu}^a+k^a+\lambda^a)\rho^b}\,
|\overline{\mu}^a+k^a-\rho^a\rangle\rangle.
\end{equation}
These actions may indeed be seen to be singled-valued under lattice shifts of the group
parameters\footnote{The composition law and cocycle properties are given hereafter.
We leave aside the construction
of an inner product on these representation spaces, as well as for those in the two
distinguished cases with $k_0=0$. This is rather straightforward. Note that in the present
case with $k_0\ne 0$, the invariant states are not normalisable since they belong to a continuous
set.} $k^a$.

Hence, in contradistinction to all other representations discussed so far, and in particular
that of the ordinary noncommutative torus in the absence of the momentum operators, the generic
irreducible representation of the noncommutative two-torus Weyl-Heisenberg group with $k_0\ne 0$ is
noncountable infinite dimensional and spanned by a collection of states labelled by two
continuous parameters each defined modulo $k_0$.

It is clear that by identifying appropriate subsets of the group parameters $(k^a,\rho^a)$,
which are closed under addition, {\it i.e.\/}, closed under composition within the NC-2T-WH group, subgroups may be
identified for which the above representation space becomes reducible, possibly leading
to discrete infinite dimensional representations of such subgroups, or even finite
dimensional ones. For instance considering only those NC-2T-WH group elements with $\rho^a=0$
the above representation space separates into an infinite noncountable ensemble of
finite $|k_0|$ dimensional representations of that subgroup. As seen from (\ref{eq:Wgen})
one then in fact constructs a representation of the subalgebra
\begin{equation}
\left[\hat{T}_a,\hat{T}_b\right]=i\hbar\,2\pi\hbar\,k_0\epsilon_{ab}\mathbb{I}
\end{equation}
of the original full noncommutative Heisenberg algebra. Since this subalgebra is
isomorphic to that of the ordinary noncommutative two-torus in Sec.~\ref{Sec3},
\begin{equation}
\left[\hat{u}^a,\hat{u}^b\right]=\frac{i}{2\pi k_0}\epsilon^{ab}\mathbb{I},
\end{equation}
and as the torus topology is defined through these operators as
translation operators, the irreducible representation of the pure
$\hat{T}_a$ algebra must indeed again be of finite dimension $|k_0|$ for some integer $k_0$.
Of course when $A=2\pi\theta k_0$ and thus $\beta=\hbar/\theta$, such a reduction
coincides precisely with the construction in Sec.~\ref{Sec3}.

In a likewise manner more involved subgroups may be imagined in which even 
nonvanishing parameters $\rho^a$ of rational values are used, but as was already
remarked at the end of Sec.~\ref{Sec2} in the commutative case, the
genuine NC-2T-WH group corresponds to all elements $W_0(k^a,\rho^a)$ for the
entire ranges of allowed values for the group parameters $(k^a,\rho^a)$.
It is thus quite remarkable that by just extending the ordinary noncommutative
configuration space algebra of operators $\hat{x}^i$ with the momentum
operators $\hat{p}_i$ on a configuration space having the topology of a torus,
the irreducible representation of finite dimension $k_0$ of the $k^2_0$
dimensional finite noncommutative Weyl-Heisenberg group
of Sec.~\ref{Sec3} turns into a noncountable infinite dimensional representation labelled
by two real variables, each defined modulo $k_0$, of a group which itself has become
the semi-direct product of a finite $k^2_0$ dimensional group and a Lie group
parametrised by the coordinates $\rho^a\in\mathbb{R}$ with specific composition
law and cocycle properties,
\begin{equation}
W_0(k^a_2,\rho^a_2)\,W_0(k^a_1,\rho^a_1)=
e^{\frac{2i\pi}{k_0}k^1_1 k^2_2+\frac{i\pi}{k_0}\epsilon_{ab}\rho^a_2\rho^b_1}\,
W_0(k^a_2+k^a_1,\rho^a_2+\rho^a_1),
\end{equation}
\begin{equation}
W_0(k^a_1,\rho^a_1)\,W_0(k^a_2,\rho^2_2)=
e^{\frac{2i\pi}{k_0}\epsilon_{ab}(\rho^a_1\rho^b_2-k^a_1 k^b_2)}\,
W_0(k^a_2,\rho^a_2)\,W_0(k^a_1,\rho^a_1).
\end{equation}

\section{The Free Particle and its Energy Spectrum}
\label{Sec7}

Given the considerations discussed in the Introduction, the choice of Hamiltonian
operator for the description of the (nonrelativistic) free particle's motion on the
noncommutative torus should be of the form
\begin{equation}
\hat{H}=\frac{1}{2}h_0\delta^{ij}\hat{\Pi}_i\hat{\Pi}_j,\qquad
h_0>0,\quad h_0\in\mathbb{R},
\end{equation}
where $\hat{\Pi}_i$ are operators built out of linear combinations of
$\hat{x}^i$ and $\hat{p}_i$ which ought to commute with the choice of
translation operators $\hat{T}_i$ in terms of which the torus lattice group
is constructed. This issue and the ensuing energy spectrum will now be
considered for each of the classes of representations addressed in the previous Sections.

\subsection{The ordinary general torus}
\label{Sec7.1}

In the ordinary commutative case with the choice of translation operators $\hat{T}_i=\hat{p}_i$,
the operators $\hat{\Pi}_i$ that commute with these are clearly the momentum operators themselves,
$\hat{\Pi}_i=\hat{p}_i$. Consequently
\begin{equation}
\hat{H}=\frac{1}{2}h_0\delta^{ij}\hat{p}_i\hat{p}_j,\qquad h_0=\frac{1}{\mu}.
\end{equation}
Since the space of invariant states is spanned by the momentum eigenstates
\begin{equation}
|\overline{m}_a\rangle,\qquad
\overline{p}_i=2\pi\hbar\tilde{e}_i^a\left(\overline{m}_a+\lambda_a\right),\qquad
\overline{m}_a\in\mathbb{Z},
\end{equation}
the eigenstates of the Hamiltonian consist precisely of these invariant states
with energy eigenspectrum
\begin{equation}
E(\overline{m}_a)=\frac{1}{2}\left(2\pi\hbar\right)^2 h_0
g^{ab}\left(\overline{m}_a+\lambda_a\right)\left(\overline{m}_b+\lambda_b\right).
\label{eq:spectrum1}
\end{equation}

\subsection{The ordinary noncommutative torus}
\label{Sec7.2}

In the case of the ordinary noncommutative algebra (\ref{eq:3}), it may readily be
established that any operator that is quadratic in the basic coordinate operators $\hat{x}^i$, and 
which commutes with the translation operators $\hat{T}_i$, which are in effect again the
$\hat{x}^i$, is necessarily proportional to the unit operator, $\mathbb{I}$.
Consequently in this situation the spectrum of the free noncommutative particle
is degenerate for each of its $k_0$ independent states for a torus area quantised in 
units of $\theta$, $A=2\pi\theta\,k_0$.

This conclusion is in accord with the fact that this specific situation is reached
as the lowest Landau level projection of the ordinary Landau problem in the absence
of any other interaction besides the coupling to the external homogeneous magnetic field.
All such states are indeed degenerate and of finite number for a torus topology of
quantised area~\cite{Connes}.

\subsection{The distinct representations with $k_0=0$}
\label{Sec7.3}

For the complete noncommutative Heisenberg algebra for which the translation operators are chosen to be
the quantities $\hat{T}_a$, defined in terms of the parameter $\beta$, we know that the
operators $\hat{Q}_a$ commute with $\hat{T}_a$, so that the general choice of Hamiltonian is
\begin{equation}
\hat{H}=\frac{1}{2}h_0 g^{ab}\hat{Q}_a\hat{Q}_b=\frac{1}{2}h_0\delta^{ij}\hat{\Pi}_i\hat{\Pi}_j,
\label{eq:Hgen}
\end{equation}
with
\begin{equation}
\hat{\Pi}_i=\tilde{e}_i^a\,\hat{Q}_a,\qquad
\hat{Q}_a=e_a^i\,\hat{\Pi}_i.
\end{equation}

When the choice $\beta=0$ is made, corresponding to $k_0=0$ with
\begin{equation}
\hat{T}_a=\hat{v}_a=\hat{Q}_a,\qquad
\hat{\Pi}_i=\hat{p}_i,
\end{equation}
the space of invariant states is spanned by the $\hat{v}_a$ eigenstates
\begin{equation}
|\overline{m}_a\rangle:\qquad \overline{v}_a=2\pi\hbar\left(\overline{m}_a+\lambda_a\right),
\qquad \overline{m}_a\in\mathbb{Z}.
\end{equation}
Consequently these states are also the energy eigenstates with energy spectrum
\begin{equation}
E(\overline{m}_a)=\frac{1}{2}\left(2\pi\hbar\right)^2 h_0 g^{ab}
\left(\overline{m}_a+\lambda_a\right)\left(\overline{m}_b+\lambda_b\right).
\end{equation}
Hence this spectrum is independent of the noncommutativity parameter $\theta$ and
in fact coincides with the one for the commutative particle.

Likewise, when the choice $\beta=2\hbar/\theta$ is made, corresponding to $k_0=0$ with
\begin{equation}
\hat{T}_a=\frac{2A\hbar}{\theta}\epsilon_{ab}\hat{U}^b=-\hat{Q}_a,\qquad
\hat{\Pi}_i=\hat{p}_i-\frac{2\hbar}{\theta}\epsilon_{ij}\hat{x}^j=-\hat{T}_i,
\end{equation}
the space of invariant states is spanned by the $\hat{U}^a$ eigenstates
\begin{equation}
|\overline{k}^a\rangle:\qquad \overline{U}^a=\frac{\pi\theta}{A}\left(\overline{k}^a+\lambda^a\right),
\qquad \overline{k}^a\in\mathbb{Z}.
\end{equation}
These are thus also the energy eigenstates of the free particle for that choice of representation,
with the energy spectrum
\begin{equation}
E(\overline{k}^a)=\frac{1}{2}\left(2\pi\hbar\right)^2 h_0 g^{ab}\left(\overline{k}^a+\lambda^a\right)
\left(\overline{k}^b+\lambda^b\right).
\end{equation}
Again this spectrum is independent of $\theta$ and coincides with the case when either $\beta=0$
or $\theta=0$.

\subsection{The generic representations with $k_0\ne 0$}
\label{Sec7.4}

In the generic situation with $k_0\ne 0$, given that the Hamiltonian is of the
form (\ref{eq:Hgen}), the relevant operators $\hat{\Pi}_i$ are
\begin{equation}
\hat{\Pi}_i=\tilde{e}_i^a\,\hat{Q}_a=
\left(1-\frac{\beta\theta}{2\hbar}\right)\hat{p}_i-\beta\epsilon_{ij}\hat{X}^j
=\hat{p}_i-\beta\epsilon_{ij}\hat{x}^j,
\end{equation}
while the translation operators are
\begin{equation}
\hat{T}_a=e_a^i
\left[\left(1-\frac{\beta\theta}{2\hbar}\right)\hat{p}_i+\beta\epsilon_{ij}\hat{X}^j\right]
=e_a^i\left[\left(1-\frac{\beta\theta}{\hbar}\right)\hat{p}_i+\beta\epsilon_{ij}\hat{x}^j\right].
\end{equation}
It then proves useful to introduce the following Fock algebra of operators,
\begin{equation}
A_i=\sqrt{\frac{A}{2\pi\hbar^2 k_0}}\left[
\beta\hat{X}^i+i\left(1-\frac{\beta\theta}{2\hbar}\right)\hat{p}_i\right],\qquad
A^\dagger_i=\sqrt{\frac{A}{2\pi\hbar^2 k_0}}\left[
\beta\hat{X}^i-i\left(1-\frac{\beta\theta}{2\hbar}\right)\hat{p}_i\right],
\end{equation}
as well as
\begin{equation}
A_\pm=\frac{1}{\sqrt{2}}\left[A_1\mp iA_2\right],\qquad
A^\dagger_\pm=\frac{1}{\sqrt{2}}\left[A^\dagger_1 \pm i A^\dagger_2\right],
\end{equation}
such that
\begin{equation}
\left[A_i,A^\dagger_j\right]=\delta_{ij}\mathbb{I},\qquad
\left[A_\pm,A^\dagger_\pm\right]=\mathbb{I}.
\end{equation}

Inverting these relations, and upon substitution into the appropriate expressions, one finds
\begin{equation}
\hat{H}=2\pi\hbar^2 k_0\frac{h_0}{A}\left[A^\dagger_+ A_++\frac{1}{2}\right],
\end{equation}
as well as
\begin{equation}
U(n^a)=e^{-i\pi k_0 n^1 n^2+2i\pi n^a\epsilon_{ab}\lambda^b}\,
e^{\sqrt{\frac{\pi k_0}{A}}\left(n^a e_a^+ A^\dagger_- - n^a e_a^- A_-\right)},
\end{equation}
where $e_a^\pm=e_a^1\pm i e_a^2$.

Considering first the NC-H algebra on the noncommutative plane rather than the two-torus,
the eigenstates of the Hamiltonian are given by
\begin{equation}
|k_+,k_-\rangle=\frac{1}{\sqrt{k_+! k_-!}}\left(A^\dagger_+\right)^{k_+}
\left(A^\dagger_-\right)^{k_-}\,|0\rangle,\qquad k_+,k_-\in\mathbb{N},
\end{equation}
$|0\rangle$ being the Fock vacuum for the $(A_\pm,A^\dagger_\pm)$ Fock algebras,
($A_\pm|0\rangle=0$) and have eigenvalues
\begin{equation}
E(k_+,k_-)=2\pi\hbar^2 k_0\frac{h_0}{A}\left[k_+ +\frac{1}{2}\right].
\label{eq:Egen}
\end{equation}
Note that this energy spectrum is once again independent of the noncommutativity parameter $\theta$.
Furthermore it is infinitely degenerate in the excitations of the $(A_-,A^\dagger_-)$ sector,
but possesses a harmonic finite gap in the excitations of the $(A_+,A^\dagger_+)$ sector,
very much like the degenerate Landau problem on the plane, the r\^ole of the
magnetic field being taken up here essentially by the integer $k_0\ne 0$, or equivalently the
parameter $\beta\ne 0,2\hbar/\theta$ according to the quantisation condition (\ref{eq:quantised}).

In order to identify now the energy eigenstates within the two-torus representation, it suffices
to apply the projection operator (\ref{eq:projector}) defined by the lattice group,
since by construction in the free particle case the Hamiltonian operator commutes with the
translation operators. Consequently the following projected energy eigenstates provide a basis for the two-torus representation:
\begin{equation}
|k_+,k_-\rangle\rangle=\mathbb{P}|k_+,k_-\rangle=
\sum_{\ell^a\in\mathbb{Z}}U(\ell^a)\,|k_+,k_-\rangle.
\end{equation}
Explicitly they read
\begin{equation}
|k_+,k_-\rangle\rangle=
\frac{1}{\sqrt{k_+! k_-!}}\left(A^\dagger_+\right)^{k_+}
\sum_{\ell^a\in\mathbb{Z}}\left[A^\dagger_- -\sqrt{\frac{\pi k_0}{A}} e_a^-\ell^a\mathbb{I}\right]^{k_-}\,
U(\ell^a)\,|0\rangle,
\end{equation}
and the energy spectrum is given in (\ref{eq:Egen}). Leaving aside the explicit construction
of a new inner product on this subspace for which these invariant energy eigenstates would be
orthonormalised\footnote{With respect to the inner product for the original orthonormalised Fock states
$|k_+,k_-\rangle$, the invariant two-torus states $|k_+,k_-\rangle\rangle$ are not normalisable.}, 
and the ensuing identification of the changes of bases $\langle\langle\overline{\mu}^a|k_+,k_-\rangle\rangle$
and $\langle\langle\overline{\nu}^a|k_+,k_-\rangle\rangle$, the important conclusion
of the above analysis is that even upon compactification onto the two-torus geometry, irrespective of
the choice of representation labelled by $k_0\ne 0$ the spectrum of the free noncommutative particle
remains totally independent of the noncommutativity parameter $\theta$.

\section{Conclusions}
\label{Sec8}

In order to identify possible observable consequences of noncommutative space coordinates in
deformations of quantum mechanical systems, the present work considered the construction of the
representations of the noncommutative Heisenberg algebra of position and momentum operators,
$\hat{x}^i$ and $\hat{p}_i$, when the configuration space topology and geometry is that of
a flat two-torus. Allowing for a general definition of the torus topology through translations
in the Euclidean configuration plane which may also transform the momentum spectrum, all possible
representations have been identified. They fall into two classes, according to whether an integer
$k_0$ labelling them is vanishing or not. When that integer $k_0$ vanishes, two distinct representations
are possible, and are essentially isomorphic to the representations of the ordinary commutative Weyl-Heisenberg algebra
on the torus spanned by a discrete spectrum of quantised momentum eigenstates and labelled by U(1) holonomy
parameters. When the integer $k_0$ is nonvanishing, translations in configuration space also shift
momentum eigenvalues, and representations of the Weyl-Heisenberg group are then continuous and spanned
by eigenstates of the momentum operators, say, of which the spectrum belongs to the fundamental domain
of some lattice structure related to the torus topology.

Note that when the configuration space translation operators are taken to be the coordinate
operators themselves, as is the case for the usual discussion of the noncommutative torus which only
considers the algebra of the position operators, a quantised torus area results.  In contrast,
by simply extending the algebra to also include the momentum operators, the representation space
changes from finite dimensional to a noncountable infinite dimensional space spanned by points
belonging to some fundamental domain.

In contrast with the single representation of the noncommutative Heisenberg algebra and
Weyl-Heisenberg group on the Euclidean plane, which is also equivalent to the commutative representation,
a rich structure of possible representations of the noncommutative Heisenberg algebra and Weyl-Heisenberg group
results on the torus. Yet, despite this rich structure, when the dynamics of a free particle is considered,
for whatever choice possible among the available representations, no physical consequence
of noncommutativity is implied. Presumably this conclusion is unavoidable in the presence of a symmetry
surviving the noncommutative deformation, namely translations in configuration space, as is also the
situation for the free particle on the noncommutative plane.

Hence, as discussed already in the Introduction, eventual observable effects of noncommutativity must
be intertwined with effects from interactions, which makes it difficult to disentangle the role
of noncommutativity and interactions on such fuzzy spaces since, at least in some approximations,
interactions may effectively be represented through noncommutativity~\cite{sch1,sch2}. The simplest
manner in which to consider interactions and still move away as little as possible from a free particle
dynamics is by confining the latter in a finite domain in configuration space through some (infinite)
well potential, in effect introducing interactions only through boundary conditions.
In the presence of noncommuting space coordinates this is not readily achieved and
a dedicated approach needs to be developed. Work on this problem is being pursued and will be reported
on elsewhere.

\section*{Acknowledgements}

J.~G. acknowledges the Institute of Theoretical Physics for an Invited Research Staff position at the
University of Stellenbosch (Republic of South Africa).
He is most grateful to Profs. Hendrik Geyer and Frederik Scholtz, and the School of Physics
for their warm and generous hospitality during his sabbatical leave, and for financial support.
His stay in South Africa is also supported in part by the Belgian National
Fund for Scientific Research (F.N.R.S.) through a travel grant.

J.~G. acknowledges the Abdus Salam International Centre for Theoretical
Physics (ICTP, Trieste, Italy) Visiting Scholar Programme
in support of a Visiting Professorship at the UNESCO-ICMPA (Republic of Benin).

The work of J.~G. is also supported by the Institut Interuniversitaire des Sciences Nucl\'eaires and by
the Belgian Federal Office for Scientific, Technical and Cultural Affairs through
the Interuniversity Attraction Poles (IAP) P6/11.

\section*{Appendix}

With respect to a choice of cartesian coordinates $x^i$ in the plane, the two-torus geometry in
characterised by lattice vectors $e_a^i$ ($a,i=1,2$) and their dual vectors $\tilde{e}_i^a$
such that
\begin{equation}
e_a^i\,\tilde{e}_i^b=\delta_a^b,\qquad
\tilde{e}_i^a\,e_a^j=\delta_i^j.
\end{equation}
The two-torus is thus defined by the equivalence relation
\begin{equation}
x^i\sim x^i+n^a\,e_a^i,\qquad n^a\in\mathbb{Z}.
\end{equation}

The torus area is given by
\begin{equation}
A=\sqrt{\det g_{ab}},\qquad g_{ab}=\delta_{ij}\,e_a^i\,e_b^j,
\end{equation}
with the inverse metric
\begin{equation}
g^{ab}=\delta^{ij}\,\tilde{e}_i^a\,\tilde{e}_j^b,\qquad
g_{ac}\,g^{cb}=\delta_a^b,\qquad
g^{ac}\,g_{cb}=\delta^a_b.
\end{equation}

The orientation of the two basis vectors $(e_1^i,e_2^i)$, in that order, is assumed
to be such that
\begin{equation}
\det e_a^i>0.
\end{equation}
Then
\begin{equation}
A=\det e_a^i,\qquad
\frac{1}{A}=\det\tilde{e}_i^a,
\end{equation}
together with
\begin{equation}
\epsilon_{ij}\,e_a^i e_b^j=A\,\epsilon_{ab},\quad
\epsilon^{ab}\,e_a^i e_b^j=A\,\epsilon^{ij},\quad
\epsilon^{ij}\,\tilde{e}_i^a\tilde{e}_j^b=\frac{1}{A}\,\epsilon^{ab},\quad
\epsilon_{ab}\,\tilde{e}_i^a\tilde{e}_j^b=\frac{1}{A}\,\epsilon_{ij},
\end{equation}
as well as
\begin{equation}
\epsilon_{ij}\,e_a^j=A\,\tilde{e}_i^b\epsilon_{ba},\quad
\epsilon_{ab}\,\tilde{e}_i^b=\frac{1}{A}\,e_a^j\epsilon_{ji},\quad
\epsilon^{ij}\,\tilde{e}_j^a=\frac{1}{A}\,e_b^i\epsilon^{ba},\quad
\epsilon^{ab}\,e_b^i=A\,\tilde{e}_j^a\epsilon^{ji},
\end{equation}
where the antisymmetric symbols $\epsilon_{ab}$ and $\epsilon_{ij}$ are
such that
\begin{equation}
\epsilon_{ij}=\epsilon^{ij},\quad
\epsilon_{ab}=\epsilon^{ab},\quad
\epsilon^{12}=+1=\epsilon_{12}.
\end{equation}

\end{document}